\documentclass[unsortedaddress,nofootinbib,10pt]{revtex4}

\usepackage[utf8]{inputenc}
\usepackage{amsmath,empheq,color}
\usepackage{amsfonts}
\usepackage{amsthm}
\usepackage{amssymb}
\usepackage{graphicx}
\usepackage{hyperref}
\usepackage[normalem]{ulem}
\usepackage{epigraph}
\usepackage{mathrsfs}
\usepackage{bbm}
\usepackage{lineno}
\usepackage{scalerel,stackengine}
\usepackage {tikz}
\usepackage{relsize}
\usepackage{extarrows}
\usepackage{caption}
\usepackage{subcaption}
\begin{document}

\title{Dirac's formalism for time-dependent Hamiltonian systems in the extended phase space}
\author{Angel Garcia-Chung}
\email{alechung@xanum.uam.mx} 
\affiliation{Departamento de F\'isica, Universidad Aut\'onoma Metropolitana - Iztapalapa, \\
San Rafael Atlixco 186, Ciudad de M\'exico 09340, M\'exico}
\affiliation{Universidad Panamericana, \\ 
Tecoyotitla 366. Col. Ex Hacienda Guadalupe Chimalistac, C.P. 01050 Ciudad de M\'exico, M\'exico}

\author{Daniel Guti\'errez-Ruiz}
 \email{daniel.gutierrez@correo.nucleares.unam.mx}

\author{J. David Vergara}
\email{vergara@nucleares.unam.mx}
\affiliation{Instituto de Ciencias Nucleares, Universidad Nacional Aut\'onoma de M\'exico, Ciudad Universitaria, Ciudad de M\'exico 04510, Mexico}

\date{\today}

\begin{abstract}
Dirac's formalism for constrained systems is applied to the analysis of time-dependent Hamiltonians in the extended phase space. We show that the Lewis invariant is a reparametrization invariant, and we calculate the Feynman propagator using the extended phase space description. We show that the Feynman propagator's quantum phase is given by the boundary term of the canonical transformation of the extended phase space. We propose a new canonical transformation within the extended phase space that leads to a Lewis invariant generalization, and we sketch some possible~applications.
\end{abstract}

\maketitle

%%%%%%%%%%%%%%%%%%%%%%%%%%%%%%%%%%%%%%%%%%%%%%%%%%%%
%%%%%%%%%%%%%%%%%%%%%%%%%%%%%%%%%%%%%%%%%%%%%%%%%%%%
%%%%%%%%%%%%%%%%%%%%%%%%%%%%%%%%%%%%%%%%%%%%%%%%%%%%
\section{Introduction}

Time-dependent Hamiltonian systems are broadly used in physics, both in classical and quantum mechanics, with many applications \cite{bayfield1999quantum, Sacha, Mukhanov}. A particular feature of these systems is that their Hamiltonians are not conserved quantities,
which makes the analytic study of the systems difficult, particularly their quantum description. However, this difficulty can be addressed using various methods, including exact invariants  \cite{mukunda1965characteristic, sudarshan1965group, lewis1969exact}. A particular class of exact invariants is the well-known Lewis invariant given in  \cite{lewis1967classical, lewis1968motion, lewis1968class} and subsequently applied to the quantum time-dependent harmonic oscillator and a charged particle in a time-dependent electromagnetic field \cite{lewis1969exact}. 

For a Hamiltonian of the form
\begin{equation}
	H(q,p,t) = \frac{p^2}{2} + \frac{\omega^2(t)\, q^2}{2}  , \label{HamilLewis}
\end{equation}
\noindent where $p$ and $q$ are conjugate canonical coordinates and $\omega(t)$ is a smooth time-dependent frequency, Lewis \cite{lewis1967classical, lewis1968motion} showed that
\begin{equation}
	I(q,p,t) = \frac{1}{2} \left( \rho \, p - \dot{\rho} \, q \right)^2 + \frac{q^2}{2 \, \rho^2}, \label{InvLewis}
\end{equation}
\noindent is an exact invariant, i.e., $\frac{d}{dt} I(q,p,t) = 0$ when the auxiliary function $\rho(t)$ satisfies
\begin{equation}
	\ddot{\rho} + \omega^2(t) \, \rho = \frac{1}{\rho^3}. \label{AuxLewis}
\end{equation}
\noindent This result can be extended to more general potentials $V(q,t)$, although only certain potentials admit such exact invariants \cite{lewis1982direct}. Leach's method in \cite{lewis1982direct} is an {\sl ad hoc} derivation that explicitly finds all invariants, either linear or quadratic, in the momentum $p$. Moreover, additional methods have been applied to associate an invariant to
time-dependent non-autonomous systems, e.g., \cite{leach1978generalization,  chattopadhyay1980noether, ray1980noether, struckmeier2000exact, struckmeier2001invariants}. 

A noteworthy example is given by Leach in \cite{leach1978generalization}, where it was shown that a two-step time-dependent linear transformation yields an exact invariant for the time-dependent harmonic oscillator. This method was later generalized by Struckmeier and Riedel \cite{struckmeier2000exact, struckmeier2001invariants}, including three-dimensional time-dependent Hamiltonians. The two-step transformation is first given by a time-dependent canonical transformation and secondly by a specific time reparametrization. The first transformation contains an arbitrary function, which is further related to the auxiliary function $\rho$ and gives rise to the auxiliary Equation (\ref{AuxLewis}). On the other hand, the time reparametrization removes the Hamiltonian's explicit time-dependence and gives rise to the Lewis invariant. In this way, time-dependent canonical transformations and time reparametrizations emerge in the analysis of exact invariants for time-dependent Hamiltonian systems.  Nonetheless, these transformations suffer restrictions in the context of time-dependent Hamiltonian mechanics \cite{abraham1978foundations}, and some caveats are in order. Let us remark on some of them.

The first restriction is that time-dependent canonical transformations ``must preserve time''; that is to say, the original and the destination system are always correlated at the same instant of their respective time scales \cite{abraham1978foundations, siegel1995lectures}.  The second restriction is that time reparametrizations are not canonical transformations \cite{abraham1978foundations, carinena1987time}. Struckmeier \cite{struckmeier2002canonical, struckmeier2005hamiltonian} solved this issue by showing the appropriate frame to study the exact invariants for a time-dependent Hamiltonian using time-dependent canonical transformations and time-reparametrizations, within the extended phase space formalism with a gauge fixed. 

The extended phase space is just an enhancement of the standard phase space where the time parameter $t$ and its conjugate momentum $p_t$ are promoted as additional canonical variables of the system.  Consequently,  the symplectic group is also enlarged, admitting in this way the time-dependent transformations of the standard phase space and time reparametrizations as particular cases of symplectic maps. 

Nevertheless, the analysis given in \cite{struckmeier2002canonical, struckmeier2005hamiltonian} by Struckmeier lacks some results that might be useful for the quantization of such systems. For example, it is not clear what commutation relation should be employed in order to canonically quantize the system or whether there is a more general  transformation to that used in \cite{struckmeier2002canonical} leading to the exact invariant of the system. These questions, among others, can be naturally answered using Dirac's method for constrained systems. As we mentioned, the enlarging of the standard phase space to the extended phase space is achieved by adding two more variables: $t$ and $p_t$. These additional variables are not really the physical ones. Dirac's method is a procedure to remove these spurious degrees of freedom consistently. 
It is broadly used in particle physics as well as in field theory physics since it provides the tools to quantize constrained systems \cite{dirac2013lectures, henneaux1992quantization}. 

Another advantage of using Dirac's method for time-dependent Hamiltonian systems in the extended phase space is that it paves the way to the path integral analysis of these systems. This is an alternative route to the known methods to determine the Feynman propagator for time-dependent Hamiltonian systems \cite{chetouani1989generalized}. These methods use the two-step transformation to modify the measure ${\cal D} q \, {\cal D}p$ in the Feynman propagator. In this case, it is possible to consider non-canonical transformations because the variables $q_j$ and $p_j$ on each infinitesimal interval are not canonically conjugated variables \cite{chetouani1989generalized}. The net effect of the first transformation is a factor depending exclusively on the auxiliary function $\rho$. On the other hand, the time reparametrization does not modify the measure ${\cal D} q \, {\cal D}p$ and contributes to the amplitude with a phase. In this derivation, it is unclear why the time reparametrization does not affect the measure ${\cal D} q \, {\cal D}p$. To answer this question, the Dirac's method can be used in the path integral description. In this case, the measure will be of the form ${\cal D}q \, {\cal D}t \, {\cal D}p \, {\cal D} p_t $, thus allowing modifications of time reparametrizations.

Regarding the extended phase space, it is worth mentioning that recent developments involving the nature of time crystals \cite{wilczek2012quantum, das2018cosmological, shapere2012classical, Sacha-2017, Autti, dai2020classical}, in which the time-translation symmetry is spontaneously broken, suggest that the extended phase space formalism might be considered as the natural arena to study these phenomena, since, in this extended phase space, the time reparametrization invariance is explicit. Additionally, other recent results in the study of quantum speed limits \cite{del2013quantum, lloyd2000ultimate, shanahan2018quantum} suggest that the extended phase space can be used to explore these limits in the context of the path integral formalism. Furthermore, in some cosmological contexts, for example, a scalar field on a Friedman-Lemaitre-Robertson-Walker (FLRW) universe, the Euler--Lagrange equations in a conformal time yield a Klein-Gordon equation with an effective time-dependent mass term \cite{Mukhanov}. This system is transformed into a tower of harmonic oscillators with time-dependent frequency once the field's Fourier decomposition is considered. Due to this time-dependence of the frequency, this system can be analyzed within the extended phase space formalism, which clarifies the quantum vacuum's construction and its connection with the Lewis invariant. Moreover, the FLRW time-invariant vacuum is built directly with the associated creation and annihilation operators, as shown in \cite{Bertoni}.

For these reasons, in addition to the analysis of the exact invariants for time-dependent Hamiltonians in the extended phase space, we consider in this work the Dirac's method for constrained systems \cite{dirac2013lectures, henneaux1992quantization} in order to fill the gaps of the analysis given by Struckmeier~\cite{struckmeier2002canonical, struckmeier2005hamiltonian}. We also obtain the Feynman propagator using the Dirac's method in the extended phase space and study the change in the measure of the path integral amplitude. Our goal is to establish a clear mathematical and physical setup in which the time-dependent Hamiltonians can be studied at both the classical and quantum levels using Dirac's formalism in the extended phase space. 

This paper is organized as follows. In Section \ref{EPSForm}, we briefly summarize the main aspects of the extended phase space analysis using Dirac's method. Section \ref{CTEPS} studies the canonical transformation and the emergence of the Lewis invariant as a Dirac observable. The Feynman propagator using the path integral analysis in the extended phase space within the Dirac's method is provided in Section \ref{PIA}. Finally, we discuss our results in Section~\ref{Discussion}.

%%%%%%%%%%%%%%%%%%%%%%%%%%%%%%%%%%%%%%%%%%%%%%%%%%%%%%%%%%%%%%%%%%%%%%%%%%%
%%%%%%%%%%%%%%%%%%%%%%%%%%%%%%%%%%%%%%%%%%%%%%%%%%%%%%%%%%%%%%%%%%%%%%%%%%%
%%%%%%%%%%%%%%%%%%%%%%%%%%%%%%%%%%%%%%%%%%%%%%%%%%%%%%%%%%%%%%%%%%%%%%%%%%%

\section{Dirac's formalism in the extended phase space} \label{EPSForm}

In this Section, we derive the main relations for time-dependent systems using Dirac's method in the extended phase space. To begin with, consider the following~action
\begin{equation}
	S\left[ q(t), \frac{dq(t)}{dt} \right] = \int^{t_2}_{t_1} \left[ \frac{m(t)}{2} \left( \frac{dq}{dt} \right)^2 - V(q,t) \right] \, dt, \label{IniAction}
\end{equation}
\noindent where the mass $m(t)$ and the time-dependent potential $V(q,t)$ are smooth functions of the time parameter $t$. As is already known, the standard Hamiltonian analysis  for the system~(\ref{IniAction}) leads to the two-dimensional phase space with coordinates $(q,p)$ and Hamiltonian function 
\begin{equation}
	H(q,p,t) := \frac{p^2}{ 2\, m(t)} + V(q,t). \label{OldHamilt}
\end{equation}
\noindent Clearly, the Hamilton equations are explicitly time-dependent 
\begin{equation}
	\frac{dq}{dt} = \frac{p}{m(t)}, \qquad \frac{dp}{dt} = - \frac{\partial V(q,t)}{\partial q} , \label{OldEqofM}
\end{equation}
\noindent hence the Hamiltonian (i.e., the energy in this case) is not a conserved quantity
\begin{equation}
	\frac{d H}{dt} = \frac{\partial H}{\partial t} = - \left( \frac{\dot{m}}{m^2} \right) \frac{p^2}{2} + \frac{\partial V(q,t)}{\partial t} \neq 0. \label{OldECLaw}
\end{equation}

Our aim is to consider the time parameter $t$ as an additional degree of freedom for the system described by (\ref{IniAction})--(\ref{OldECLaw}). To do so, we consider the arbitrary time reparametrization $t=\tilde{t}(\tau)$, where the parameter $\tau$ plays the role of the new time parameter (a similar analysis but that can be found in a cosmological context in \cite{maeder2017alternative}). The function $\tilde{t}(\tau)$ is  chosen so as to give a smooth one-to-one correspondence of the domains of $\tau$ and $t$ (see, for instance,~\cite{fulop1999reparametrization,deriglasov}). This transformation changes the dependence of the coordinate function $q(t)$; thus, the following definition $q(\tilde{t}(\tau)) =: \tilde{q}(\tau) $ is required. In this scheme, $t$ is no longer the time but an additional dynamical variable of the system and, consequently, a new functional in the expression for the action $S$ in (\ref{IniAction}), which we denote now by $\tilde{S}$ 
\begin{equation} 
	\tilde{S}\left[  \tilde{q}, \tilde{t}, \frac{d \tilde{q}}{d\tau}, \frac{d \tilde{t}}{d\tau} \right] = \int^{\tau_2}_{\tau_1} \left[ \frac{m(\tilde{t})}{2} \left( \frac{\tilde{q}'}{\tilde{t}'}\right)^2 - V(\tilde{q},\tilde{t}) \right] \tilde{t}' \, d\tau, \label{GAction}
\end{equation}
\noindent where the prime means derivative with respect to $\tau$. Note that both $\tilde{q}$ and $\tilde{t}$ constitute the generalized configuration variables on this extended space, whereas $\tilde{q}'=d \tilde{q}/d\tau$ and $\tilde{t}'=d \tilde{t}/d\tau$ correspond to their generalized velocities.  At this point, it is worth  mentioning the following remarks: (i) in cases in which we are considering a relativistic particle, this reparametrization procedure is also applicable, and (ii) a relativistic particle is already a reparametrization-invariant system. More details can be found in \cite{LucioNiertoDavid, israel1973relativity}.

If we consider a new reparametrization $\tau =\tilde{\tau}(\sigma)$, it gives rise to the following relation
\begin{equation} \label{Symmetry}
	\tilde{S}\left[ \tilde{q}, \tilde{t}, \frac{d \tilde{q}}{d\tau},  \frac{d \tilde{t}}{d\tau} \right] = \tilde{S}\left[ \tilde{\tilde{q}}, \tilde{\tilde{t}}, \frac{d \tilde{\tilde{q}}}{d\sigma} ,  \frac{d \tilde{\tilde{t}}}{d\sigma} \right],
\end{equation}
\noindent where $ \tilde{\tilde{q}}(\sigma) := \tilde{q}(\tilde{\tau}(\sigma))$ and $ \tilde{\tilde{t}}(\sigma) := \tilde{t}(\tilde{\tau}(\sigma))$. This shows that the action $\tilde{S}$ is invariant under reparametrizations \cite{fulop1999reparametrization}. Naturally, the emergence of this symmetry (\ref{Symmetry}) is a direct consequence of the enlarging of the dynamical degrees of freedom, and it will affect the Hamiltonian analysis as we will see further. 

Consider now the Hessian matrix for (\ref{GAction})
\begin{equation}
	{\cal H} = \left( \begin{array}{cc} \frac{\partial^2 \tilde{S}}{\partial \tilde{q}'^2} & \frac{\partial^2 \tilde{S}}{\partial \tilde{q}' \partial \tilde{t}'} \\ \frac{\partial^2 \tilde{S}}{\partial \tilde{t}' \partial \tilde{q}'} & \frac{\partial^2 \tilde{S}}{\partial \tilde{t}'^2} \end{array}\right) = \frac{m(\tilde{t})}{\tilde{t}'^3} \left( \begin{array}{cc} \tilde{t}'^2 & - \tilde{q}' \tilde{t}' \\ - \tilde{q}' \tilde{t}' & \tilde{q}'^2 \end{array}\right), \label{Hessian}
\end{equation}
\noindent and whose determinant is equal to zero. This Hessian matrix (\ref{Hessian}) is not invertible, and consequently, the Euler--Lagrange equation for $\tilde{t}$ is not independent of the Euler--Lagrange equation for $\tilde{q}$. This implies that the extended system in the variables $\tilde{t}$ and $\tilde{q}$ is a constrained system \cite{dirac2013lectures,henneaux1992quantization}; therefore, Dirac's formalism can be applied to the action (\ref{GAction}). 

To do so, let us first consider the generalized canonical conjugate momenta
\begin{eqnarray}
	p &=& \frac{\delta S}{\delta \tilde{q}'} = \frac{m(\tilde{t}) \, \tilde{q}'}{\tilde{t}'}, \label{CMp}\\
	p_{\tilde{t}} &=& \frac{\delta S}{\delta \tilde{t}'} = - \left[ \frac{m(\tilde{t})}{2} \frac{\tilde{q}'^2}{\tilde{t}'^2} + V(\tilde{q},\tilde{t}) \right]. \label{TMomentum}
\end{eqnarray}
\noindent Using Equation (\ref{CMp}), we can express $\tilde{q}'$ in terms of the momentum $p$ as
\begin{equation} 
	\tilde{q}' = \frac{ \tilde{t}' }{m(\tilde{t})} \, p,
\end{equation}
\noindent and inserting the former relation in the expression for $p_{\tilde{t}}$ given in (\ref{TMomentum}), we obtain the primary~constraint
\begin{equation} \label{Constraint}
	\phi = p_{\tilde{t}} + H(\tilde{q},p,\tilde{t}) \approx 0 , 
\end{equation}
\noindent where $H(\tilde{q},p,\tilde{t}) $ is the Hamiltonian given in (\ref{OldHamilt}). Let us emphasize that this is the general expression of the constraint; thus, it can be used for any time-dependent mass $m(t)$ term or potential $V(q,t)$ (see, for example, \cite{Mukhanov}). Moreover, once we consider $p = \partial S/\partial q$ and $p_t = \partial S / \partial t $, this constraint gives rise to the Hamilton--Jacobi equation---more details \mbox{in~\cite{van2014time, fanchi1993parametrized, guler1992canonical, dominici1984hamilton, pimentel1998hamilton, rothe2003hamilton}}.

The weak equality symbol $\approx$ stands for the following: the Poisson bracket $\{ B_1, B_2 \}$ of two arbitrary functions $B_1(\tilde{q}, \tilde{t},p,p_{\tilde{t}})$ and $B_2(\tilde{q}, \tilde{t},p,p_{\tilde{t}})$ on the extended phase space evaluated on the constraint surface $\phi =0$ are different from that of the Poisson bracket of the same functions but first evaluated on the constraint $B_1(\tilde{q}, \tilde{t},p,-H)$ and  $B_2(\tilde{q}, \tilde{t},p,-H)$; that is to say
\begin{equation}
	\{ B_1, B_2 \}_{\phi=0} \neq \{ {B_1}_{\phi=0}, {B_2}_{\phi=0} \} .
\end{equation}

To avoid this contradiction, Dirac proposed to take \cite{dirac2013lectures}, as a rule, to first work out the Poisson brackets before we make use of the constraint equation $\phi=0$. Therefore, the weak equality sign reminds us of this rule \cite{dirac2013lectures, henneaux1992quantization}.  

The next step is to consider the canonical Hamiltonian, which takes the form
\begin{equation}
	H_c = p \tilde{q}' + p_{\tilde{t}} \tilde{t}' - L = \tilde{t}' \phi , 
\end{equation}
\noindent where
\begin{equation}
	L:= \left[ \frac{m(\tilde{t})}{2} \left( \frac{\tilde{q}'}{\tilde{t}'}\right)^2 - V(\tilde{q},\tilde{t}) \right] \tilde{t}' .
\end{equation} 
\noindent Hence, the total Hamiltonian $H_T$ is proportional to the constraint $\phi$, and the coefficient of proportionality $\lambda$ is the Lagrange multiplier. This Lagrange multiplier is a function of the parameter $\tau$, and it is independent of the phase space points. This type of Lagrange multiplier is referred to as non-canonical gauge \cite{henneaux1992quantization}. In other words, the Hamiltonian $H_T$ can be written as  
\begin{equation}
	H_T = \lambda \phi,
\end{equation}
\noindent Hence, the total Hamiltonian is null when the constraint is strongly zero. Systems with this particular type of Hamiltonian are usually called reparametrization-invariant systems and can be found in many physical scenarios, e.g., the free relativistic particle, the FLRW universe, and the canonical formulation of the General Relativity \cite{fulop1999reparametrization}.

The Hamilton equations 
\begin{equation}
	\tilde{q}' = \lambda \, \frac{p}{m(\tilde{t})}, \qquad p' = - \lambda \frac{\partial V}{\partial \tilde{q}}, \qquad \tilde{t}' = \lambda , \qquad p'_{\tilde{t}} = - \lambda \frac{\partial H}{\partial \tilde{t}} = \lambda \left[ \frac{\dot{m}}{m^2} \frac{p^2}{2} - \frac{\partial V}{\partial \tilde{t}}\right] , \label{FHEq}
\end{equation}
\noindent can be derived using the Poisson brackets
\begin{equation}
	\{ B_1, B_2  \}_{PB} = \frac{\partial B_1}{\partial \tilde{q}} \frac{\partial B_2}{\partial p} + \frac{\partial B_1}{\partial \tilde{t}} \frac{\partial B_2}{\partial p_{\tilde{t}}} - \frac{\partial B_1}{\partial p} \frac{\partial B_2}{\partial \tilde{q}} - \frac{\partial B_1}{\partial p_{\tilde{t}}}\frac{\partial B_2}{\partial \tilde{t}}, \label{PB}
\end{equation}
\noindent where $B_1$ and $B_2$ are two arbitrary smooth functions over the extended phase space. Here, the $\tau$-evolution is given by the standard expression $ A' = \left\{ A , H_T \right\}_{PB}$ and the Hamiltonian $H_T$. Note that, as expected, there is no equation for the Lagrange multiplier $\lambda$, and consequently, two different Lagrange multipliers induce a different evolution of the same initial point on the extended phase space. 

The process in which a value for the Lagrange multiplier $\lambda$ is fixed is usually called ``fixing the gauge''. For instance, the most common gauge fixing is the case in which $\lambda =1$. In this case, the gauge solves the third equation in (\ref{FHEq}), whereas the first two equations give rise to the Hamilton Equation (\ref{OldEqofM}) and turn the fourth equation into  (\ref{OldECLaw}) once we make the following identification $p'_{t} = - dH/dt$. As a result, we recover the system given by (\ref{OldEqofM}) with Hamiltonian (\ref{OldHamilt}) and energy evolution (\ref{OldECLaw}). However, this procedure, although quite direct and simple, does not show the Poisson bracket structure for the resulting physical degrees of freedom that might appear, for instance, in more complicated systems.

To do so, the gauge fixing condition is promoted as an additional constraint surface $\eta \approx 0$ such that $\{ \phi, \eta \} \not \approx 0.$ Under this premise, the Poisson bracket (\ref{PB}) is replaced by the so-called Dirac bracket, which, in this case, is given by
\begin{equation}
	\{ B_1, B_2 \}_{DB} = \{ B_1, B_2 \}_{PB} + \frac{1}{ \{ \phi, \eta \}_{PB} } \left( \{ B_1, \phi \}_{PB} \{ \eta, B_2\}_{PB} - \{ B_1, \eta \}_{PB} \{ \phi, B_2 \}_{PB} \right). \label{DB}
\end{equation}

Note that the constraint $\eta = \tilde{t} - \tau \approx 0$ is tantamount to fixing the gauge $\lambda =1$, and it gives the following non-null Dirac brackets 
\begin{equation} \label{DBra}
	\{ \tilde{q}, p \}_{DB} =1, \qquad \{ \tilde{q}, p_{\tilde{t}} \}_{DB} = - \frac{\partial \phi}{\partial p} = - \frac{p}{m}, \qquad \{ p, p_{\tilde{t}} \}_{DB} = \frac{\partial \phi}{\partial \tilde{q}} = \frac{\partial V}{\partial \tilde{q}}.
\end{equation} 
\noindent It can be seen that the last two relations differ from their counterparts in terms of the Poisson brackets $\{ \tilde{q}, p_{\tilde{t}} \} _{PB} = \{ p, p_{\tilde{t}} \} _{PB}  =0 $. Notably, the Dirac bracket $\{ \tilde{t}, p_{\tilde{t}} \}_{DB} = 0$ differs from its Poisson bracket pair $\{ \tilde{t}, p_{\tilde{t}} \}_{PB} = 1$. In this way,  adding another constraint $\eta \approx 0$ allows us to define a reduced phase space, which is the subspace of the phase space $(\tilde{q}, \tilde{t}, p, p_{\tilde{t}} )$ such that the constraints are satisfied; i.e., $ \phi=0 $ and $ \eta=0 $. It is in this reduced phase space where the dynamic of the system takes place, and it is generated by the Hamiltonian $H$ together with the Dirac brackets (\ref{DBra}). To see this, we first select the physical degrees of freedom, which, in this case, due to the constraints $\tilde{t} = \tau$ and $p_{\tilde{t}} = - H(\tau,\tilde{q},p)$, yield the coordinates $\tilde{q}$ and $p$ as the physical degrees of freedom in accordance with the previous~description. 

To conclude this brief analysis, consider the Hamiltonian form of the action (\ref{GAction}) 
\begin{equation}
	\tilde{S} = \int^{\tau_2}_{\tau_1} \left[ p \tilde{q}' + p_{\tilde{t}} \tilde{t}' - \lambda \phi  \right] \, d\tau, \label{AcHamil}
\end{equation}
\noindent and let us evaluate (\ref{AcHamil}) on the constraints $\phi=0$ and $\eta=0$. As a result, we will obtain the~action
\begin{equation}
	\tilde{S} = \int^{t_2}_{t_1} \left[ p \tilde{q}'  - H(\tilde{q},p,t)  \right] \, dt, \label{HAction}
\end{equation}
\noindent which is the standard Hamiltonian form of the action (\ref{IniAction}).

To summarize, we have shown that the Dirac's formalism for the system given by the action (\ref{IniAction}) in the extended phase space gives rise to the standard Hamiltonian analysis with Hamiltonian $H$ and coordinates $(\tilde{q},p)$. In this formalism, the time variable $t$ is promoted as an additional dynamical variable $\tilde{t}$. Its conjugate momentum results from the consideration of the extended action given in (\ref{GAction}). The converse procedure, namely removing these extra degrees of freedom, is accomplished by the gauge fixing process $\lambda=1$ and by the introduction of an additional constraint $\eta \approx 0$. These steps are a result of the implementation of Dirac's scheme. 

It is worth  mentioning that the gauge fixing process is mainly used within the path integral formulation of quantum mechanics, while the additional constraint procedure is commonly used in the canonical quantization scheme in order to derive the Dirac brackets and the reduced phase space.

As we mentioned, a direct implication of the aforementioned enlarging of the phase space is that it also enlarges the system's symplectic group. With this in mind, in the following subsection, we  study a canonical transformation of the extended phase space such that the final dynamical description of the reduced phase space is no longer time-dependent. To achieve this, we  use the ``extended'' energy conservation law as an auxiliary equation for the canonical map's arbitrary coefficient. This canonical transformation is a generalization of the Struckmeier transformation given in \cite{struckmeier2001invariants}. Moreover, we  show that this transformation gives rise to a boundary term that can be related to the Lewis phase in the path integral quantization.

\section{Canonical transformation in the extended phase space} \label{CTEPS}

Recall that a canonical transformation in the symplectic space $(\Gamma, \omega)$ is a map ${\cal M} : \Gamma \rightarrow \Gamma$ such that the symplectic two-form $\omega$ is preserved \cite{siegel1995lectures}. This definition is mathematically expressed as
\begin{equation}
{\cal M}^T \, J \, {\cal M} = J, \label{CTCond}
\end{equation}
\noindent where $J$ is the complex structure map $J^2 = -1$, which in matrix form reads as
\begin{equation}
J = \left( \begin{array}{cc} 0 & 1 \\ -1 & 0 \end{array}\right).
\end{equation}
 
The Leach-Struckmeier transformation \cite{leach1978generalization, struckmeier2001invariants} consists of two transformations in the standard phase space with coordinates $(q, p)$, one of which is a canonical transformation and the second is a time reparametrization, which is not a canonical transformation \cite{carinena1987time}. However, as a result of the enlarging of the symplectic group, the time reparametrization can be considered as a canonical transformation in the extended phase space. 

Our goal is to consider the more general canonical transformation in the extended phase space containing the Struckmeier transformation. Of course, in this case, time reparametrization can be considered as a part of the canonical transformation together with a given variation for the variable $p_{\tilde{t}}$. Let us consider a coordinate transformation of the form 
\begin{equation}
\left( \begin{array}{c} \tilde{q} \\ \tilde{t} \\ p \\ p_{\tilde{t}} \end{array}\right) = \left( \begin{array}{c} A(Q,T) \\ B(Q,T) \\ C(Q,T) P + D(Q,T)  \\ G(Q,T,P, P_T)\end{array}\right), \label{PSCTrans}
\end{equation}
where the coefficients $A(Q,T)$, $B(Q,T)$, $C(Q,T)$, and $D(Q,T)$, together with $G(Q, T, P, P_T)$, are smooth functions to be determined in order to make (\ref{PSCTrans}) canonical. We take the variables $\tilde{q}$ and $\tilde{t}$ to be momenta-independent to avoid a more sophisticated dependence of the new function $K(Q,P,T):= H(\tilde{q}(Q,T),\tilde{t}(Q,T),p(Q,T,P))$ in terms of the new momentum $P$.  For the same reason, we select a linear relation between the momenta $p$ and $P$ to preserve the square order of the kinetic term on $K$. We remark that, on the extended phase space, the canonical transformation is ``time-independent'', since it does not have $\tau$ dependence. Furthermore, note that the momentum $p$ is also $P_T$-independent to avoid a term of the form $P^2_T$, which, after the gauge-fixing procedure, gives rise to negative energy solutions. The canonical transformation matrix for (\ref{PSCTrans}) is given as
\begin{equation}
{\cal M} = \left( \begin{array}{cccc} \frac{\partial \tilde{q}}{\partial Q} & \frac{\partial \tilde{q}}{\partial T} & \frac{\partial \tilde{q}}{\partial P} & \frac{\partial \tilde{q}}{\partial P_T} \\ \frac{\partial \tilde{t}}{\partial Q} & \frac{\partial \tilde{t}}{\partial T} & \frac{\partial \tilde{t}}{\partial P} & \frac{\partial \tilde{t}}{\partial P_T} \\ \frac{\partial p}{\partial Q} & \frac{\partial p}{\partial T} & \frac{\partial p}{\partial P} & \frac{\partial p}{\partial P_T} \\ \frac{\partial p_t}{\partial Q} & \frac{\partial p_t}{\partial T} & \frac{\partial p_t}{\partial P} & \frac{\partial p_t}{\partial P_T} \end{array}\right) = \left( \begin{array}{cccc} A' & \dot{A}  & 0 & 0 \\ B' & \dot{B} & 0 & 0 \\ C' P+D' & \dot{C} P + \dot{D}  & C & 0 \\ \frac{\partial G}{\partial Q} & \frac{\partial G}{\partial T} & \frac{\partial G}{\partial P} & \frac{\partial G}{\partial P_T} \end{array}\right) . \label{JMExt}
\end{equation}

Henceforth, the prime and the dot symbols denote the operators $\frac{\partial}{\partial Q} $ and $\frac{\partial}{\partial T}$, respectively. In order to make (\ref{PSCTrans}) canonical, ${\cal M}$ must satisfy the condition (\ref{CTCond}). This yields the following system of partial differential equations 
\begin{eqnarray}
\dot{A} (C' \, P + D' ) + \dot{B} G' &=& A' (\dot{C} P + \dot{D}) + B' \dot{G} , \\
\dot{A} \, C + \dot{B} \, \frac{\partial G}{\partial P} &=& 0, \\
B' \, \frac{\partial G}{\partial P} + A' \, C &=& 1 , \\
\dot{B}\, \frac{\partial G}{\partial P_T} &=& 1, \\
B' \, \frac{\partial G}{\partial P_T}  &=& 0,
\end{eqnarray}
\noindent whose general solution is
\begin{equation}
\tilde{t} = B = B(T), \qquad C = \frac{1}{A'}, \qquad p_{\tilde{t}} = G= \frac{P_T}{\dot{B}} - \frac{\dot{A}\, P}{A' \, \dot{B}} + \frac{1}{\dot{B}} \int \left( A' \, \dot{D} - \dot{A} \, D' \right) dQ .
\end{equation}
Consequently, the new coordinates $(\tilde{q}, \tilde{t}, p, p_t)$ can be written in terms of the old coordinates as
\begin{equation}
\left( \begin{array}{c} \tilde{q} \\ \tilde{t} \\ p \\ p_{\tilde{t}} \end{array}\right) = \left( \begin{array}{c} A(Q,T) \\ B(T) \\ \frac{1}{A'} P + D(Q,T)  \\ \frac{P_T}{\dot{B}} - \frac{\dot{A}\, P}{A' \, \dot{B}} + \frac{1}{\dot{B}} \int \left( A' \, \dot{D} - \dot{A} \, D' \right) dQ  \end{array}\right). \label{CTEPhaseS}
\end{equation}

Remarkably, the momentum $p_{\tilde{t}}$ results to be a linear function in terms of the momenta $P_T$ and $P$, i.e., a contact transformation. On the other hand, note that the functions $B(T)$, $A(Q,T)$ and $D(Q,T)$ are arbitrary so far. We will see further that additional conditions to the new Hamiltonian are necessary to remove its time-dependence and that these conditions can be used to fix these coefficients via some differential equations. 

The constraint $\phi$ can be written in terms of the new variables after inserting (\ref{CTEPhaseS}) into (\ref{Constraint})
\begin{equation}
\phi = \frac{1}{\dot{B}} \left\{ P_T - \left[ \left( - \frac{\dot{B}\, P^2 }{2 \, m \, A'^2} \right) + \left( \frac{\dot{A}}{A'} - \frac{\dot{B} \, D}{m \, A'} \right) P - \int \left( A' \dot{D} - \dot{A} D' \right) dQ - \frac{\dot{B} \, D^2}{2\, m} - \dot{B} \, V \right] \right\}. \label{ConstraintNV}
\end{equation}

As can be seen, the constraint is naturally split into two parts. The first is the linear term $P_T$ and the second one is the expression inside the square brackets. The linear term is connected with the fact that $p_{\tilde{t}}$ is linear in $P_T$ and that $p$ is $P_T$-independent. Furthermore, due to this linear relation, we select the new momentum $P_T$ as a non-physical degree of freedom. The second term in (\ref{ConstraintNV}), the one inside the square brackets, will contribute to the new Hamiltonian function and to a boundary term associated with this canonical transformation.

The equations of motion in these variables take the form
\begin{eqnarray}
&& \frac{d Q}{d \tau} = \{ Q, \lambda \phi \} =  \frac{\lambda}{\dot{B}} \left[ \left( \frac{\dot{B}}{m \, A'^2} \right) P + \left( \frac{\dot{B} \, D}{m \, A'} - \frac{\dot{A}}{A'} \right) \right] , \\
&& \frac{d P}{d \tau} = \{ P, \lambda \phi \} =  - \frac{\lambda}{\dot{B}} \left[ \frac{\partial}{\partial Q} \left( \frac{\dot{B}}{2m\, A'^2} \right) P^2 + \frac{\partial}{\partial Q} \left( \frac{\dot{B} \, D}{m \, A'} - \frac{\dot{A}}{A'}\right) P + \left( A' \dot{D} - \dot{A} D' \right) + \frac{\partial}{\partial Q} \left( \frac{\dot{B} D^2}{2m} + \dot{B} V \right)\right] , \\
&& \frac{d T}{d \tau} = \{ T, \lambda \phi \} =  \frac{\lambda}{\dot{B}},\label{EqforT}\\
&& \frac{d P_T}{d \tau} = \{ P_T, \lambda \phi \} =- \frac{\lambda}{\dot{B}} \left\{  \frac{\partial}{\partial T} \left( \frac{\dot{B}}{2m\, A'^2} \right) P^2 + \frac{\partial}{\partial T} \left( \frac{\dot{B} \, D}{m \, A'} - \frac{\dot{A}}{A'}\right) P + \frac{\partial}{\partial T} \left[ \int dQ (A' \dot{D} - \dot{A} D' )+ \frac{\dot{B} D^2}{2m} + \dot{B} V \right] \right\} . \nonumber \\
\end{eqnarray}

In the last equation, we made the constraint strongly equal to zero; thus, the term proportional to the constraint $- \lambda \frac{\partial}{\partial T} \left( \frac{1}{\dot{B}}\right) \phi $ is removed from the final expression. We now look upon the gauge fixing process and take $\frac{d T}{d \tau } =1$, or equivalently,  $\eta = T - \tau \approx 0$, in which case the previous equations read as
\begin{eqnarray}
&& \frac{d Q}{d T} =  \left( \frac{\dot{B}}{m \, A'^2} \right) P + \left( \frac{\dot{B} \, D}{m \, A'} - \frac{\dot{A}}{A'} \right) , \label{EqofMQ} \\
&& \frac{d P}{d T} = - \left[ \frac{\partial}{\partial Q} \left( \frac{\dot{B}}{2m\, A'^2} \right) P^2 + \frac{\partial}{\partial Q} \left( \frac{\dot{B} \, D}{m \, A'} - \frac{\dot{A}}{A'}\right) P + \left( A' \dot{D} - \dot{A} D' \right) + \frac{\partial}{\partial Q} \left( \frac{\dot{B} D^2}{2m} + \dot{B} V \right)\right] , \label{EqofMP} \\
&& \frac{d P_T}{d T} = - \left\{  \frac{\partial}{\partial T} \left( \frac{\dot{B}}{2m\, A'^2} \right) P^2 + \frac{\partial}{\partial T} \left( \frac{\dot{B} \, D}{m \, A'} - \frac{\dot{A}}{A'}\right) P + \frac{\partial}{\partial T} \left[ \int dQ (A' \dot{D} - \dot{A} D' )+ \frac{\dot{B} D^2}{2m} + \dot{B} V \right] \right\} . \label{ECLaw}
\end{eqnarray}

The first two equations are the Hamilton equations once $Q$ and $P$ are fixed as the physical degrees of freedom. Consequently, the last equation provides the energy conservation law for this system in the coordinates $(Q,P)$. As we already know, the system in the reduced phase space is not conservative due to $H(\tilde{q},p,\tilde{t})$ being explicitly time-dependent (see Equation (\ref{FHEq})). Nevertheless, using this last Equation (\ref{ECLaw}), the system will be a conservative system if and only if
\begin{equation}
\frac{d P_T}{d T} =0.
\end{equation}

This condition must hold in the full reduced phase space $(Q,P)$; hence, the right-hand side of (\ref{ECLaw}) results in the following equations
\begin{eqnarray}
\frac{\partial}{\partial T} \left( \frac{\dot{B}}{2m\, A'^2} \right) =0, \qquad  \frac{\partial}{\partial T} \left( \frac{\dot{B} \, D}{m \, A'} - \frac{\dot{A}}{A'}\right) =0, \qquad  \frac{\partial}{\partial T} \left[ \int dQ (A' \dot{D} - \dot{A} D' )+ \frac{\dot{B} D^2}{2m} + \dot{B} V \right] =0.  \label{EnergyCL}
\end{eqnarray}

The first equation can be easily solved for $\dot{B}$ as
\begin{equation}
\frac{ \partial B}{\partial T} = \frac{m(B(T))}{m_0} \left( \frac{ \partial A}{\partial Q}  \right)^2, \label{BDot}
\end{equation}
\noindent where $1/m_0$ is the integration constant. Due to $B$ being $Q$-independent, then $A(Q,T)$ must be of the form
\begin{equation}
A(Q,T) = \tilde{A}(T) \, Q, \label{AinQ}  
\end{equation}
\noindent where $\tilde{A}(T)$ is a smooth function of $T$ to be fixed, and we assume that the integration constant $m_0$ is $Q$-independent. Of course, a more general solution implies considering $m_0 = m_0(Q)$, but this results in a $Q$-dependent mass term for the new Hamiltonian, which exceeds the purpose of the present work. Combining (\ref{BDot}) and (\ref{AinQ}), we obtain that 
\begin{equation}
t = B(T) := \tilde{B}^{-1}\left(\frac{1}{m_0} \int \tilde{A}^2(T) \, dT \right) , \qquad \mbox{where} \quad \tilde{B}(t) := \int \frac{dt}{m(t)}.  \label{ExBandD1}
\end{equation}

Inserting  (\ref{BDot}), (\ref{AinQ}), and (\ref{ExBandD1}) into the second equation of (\ref{EnergyCL}), we obtain the expression for $D(Q,T)$ in terms of $\tilde{A}(T)$ as
\begin{equation}
	D(Q,T) = \frac{m_0}{\tilde{A}} \left[ \kappa(Q) + \frac{\dot{\tilde{A}}}{\tilde{A}} Q \right] , \label{ExBandD2}
\end{equation}
\noindent where $\kappa(Q)$ is the integration function $T$-independent of the second equation in (\ref{EnergyCL}).

In order to fix $\tilde{A}(T)$, we use the third equation in (\ref{EnergyCL}) which, after inserting the former definitions, takes the form
\begin{equation}
	\frac{\partial}{\partial T} \left[ \frac{m_0 \, Q^2}{2} \left( \frac{\ddot{\tilde{A}}}{\tilde{A}} -  \frac{2\,\dot{\tilde{A}}^2}{\tilde{A}^2} \right)+ \frac{m(T) \, \tilde{A}^2 }{m_0} \, V(\tilde{q}(Q,T),\tilde{t}(T)) \right]  = 0. \label{EqforA}
\end{equation}

This is the equation for $\tilde{A}(T)$ in terms of $m(T)$ and $V(\tilde{q}(Q,T),\tilde{t}(T))$. As can be seen, for a general potential $V(A(Q,T),B(T))$, this equation is not necessarily $Q$-independent. This is an inconvenient result implying that only potentials of the form $V(\tilde{q},\tilde{t}) \sim \tilde{q}^2 $ give rise to an Equation (\ref{EqforA}) independent of the coordinate $Q$ \cite{struckmeier2001invariants}. The term proportional to $Q^2$ is the contribution of the canonical transformation in addition to the term given by the potential $V(\tilde{q}(Q,T),\tilde{t}(T))$. To solve Equation (\ref{EqforA}) for an arbitrary potential $V(\tilde{q},\tilde{t})$, the  Equations (\ref{EqofMQ}) and (\ref{EqofMP}) have to be considered\footnote{An alternative analysis is to consider $m_0$ as a function of $Q$ and such that the $Q$-dependence is removed from (\ref{EqforA}). Again, this can only work for particular cases of the potential $V$.}. 

Once we obtain the expression for $\tilde{A}(T)$, we integrate (\ref{EqforA}) and use (\ref{ExBandD1}) and (\ref{ExBandD2}) to~obtain
\begin{equation}
\frac{m_0 \, \kappa(Q)^2}{2} +  \frac{m_0 \, Q^2}{2} \left( \frac{\ddot{\tilde{A}}}{\tilde{A}} -  \frac{2\,\dot{\tilde{A}}^2}{\tilde{A}^2} \right)+ \frac{m(T) \, \tilde{A}^2}{m_0} \, V(\tilde{q}(Q,T),\tilde{t}(T)) =: \overline{V}(Q), \label{DefNV}
\end{equation}
\noindent where the function $\overline{V}(Q)$ is the time-independent potential for the physical degrees of freedom $(Q,P)$. We can see that, by inserting (\ref{AinQ}), (\ref{ExBandD1}), (\ref{ExBandD2}), and (\ref{DefNV}) into (\ref{EqofMQ}) and (\ref{EqofMP}), this potential leads to the following Hamilton equations
\begin{equation} \label{NewHEq}
\frac{d Q}{d T} =  \frac{1}{m_0} P + \kappa(Q) , \qquad  \frac{d P}{d T} = - \frac{\partial \kappa(Q)}{\partial Q} P -  \frac{\partial \overline{V}(Q)}{\partial Q} .
\end{equation}

In order to derive the Hamiltonian function giving rise to these Hamilton equations, let us consider the Hamiltonian form of the action (\ref{AcHamil}) in the new coordinates
\begin{equation}
S = \int^{\tau_2}_{\tau_1} \left\{  \left( P  + A' \, D \right)  \frac{d Q}{d \tau } + \left[ P_T + \dot{A} \, D +  \int \left( A' \, \dot{D} - \dot{A} \, D' \right) dQ \right] \frac{d T}{d \tau } - \lambda \phi \right\} d\tau, \label{ActionNV}
\end{equation}
\noindent where $\phi$ is given in (\ref{ConstraintNV}), and recall that the Lagrange multiplier given by Equation (\ref{EqforT}) is $\lambda = \dot{B}(\tau)$. Solving the constraint (\ref{ConstraintNV}) for the momentum $P_T$ and inserting the result into~(\ref{ActionNV})~gives
\begin{equation}
S =  \int^{T_2}_{T_1} \left\{  P \frac{d Q}{d T } -I(Q,P) \right\} d\, T + \left[ \int (A' \, D) dQ \right]^{T_2}_{T_1},  \label{NewS}
\end{equation}
\noindent where  
 \begin{equation}
I(Q,P):= \frac{P^2 }{2 \, m_0 }  + \kappa(Q) P  + \overline{V}(Q), \label{NEWHamil}
\end{equation}
\noindent is the new Hamiltonian and the last term in (\ref{NewS}) is the boundary term associated with the canonical transformation.

The boundary term given in (\ref{NewS}) will play a major role in the quantum description of the system, particularly in the path integral analysis. The Hamiltonian (\ref{NEWHamil}), on the other hand, is the Lewis invariant given by Struckmeier et al. \cite{struckmeier2001invariants} when it is written in terms of the coordinates $(q,p,t)$; of course, $I(Q,P)$ is clearly time-independent. Moreover, $I(Q,P)$ is a gauge-invariant observable; that is to say, it commutes with the constraint $\{ I, \phi \}_{PB} = 0$. The significance of this result is that $I(Q,P)$ is not only time-independent but also gauge-independent. Due to the gauge symmetry in this system, there is a time-reparametrization symmetry, and thus the Lewis invariant is nothing but a reparametrization invariant: {\it we can use a different gauge for $\lambda$ and the result will be the same $I(Q,P)$}.

The Dirac brackets in the variables $Q, T, P, P_T$ can be obtained as
\begin{equation}
\{ B_1, B_2 \}_{DB} := \{ B_1, B_2 \}_{PB} + \frac{1}{ \{ \phi, \eta \}_{PB} } \left(  \{ B_1, \phi \}_{PB}  \{ \eta, B_2 \}_{PB} -  \{ B_1, \phi \}_{PB}  \{ \eta, B_2 \}_{PB} \right),
\end{equation}
\noindent where $\eta := T - \tau \approx 0 $ and $\phi$ is given in (\ref{ConstraintNV}). Note that $\{ \phi , \eta \}_{PB} = -1/\dot{B}$, and as a result, the only non-null Dirac brackets are
\begin{equation}
\{ Q, P \}_{DB} = 1, \qquad \{ Q, P_T \}_{DB} = - \dot{B} \left( \frac{\partial \phi}{\partial P} \right), \qquad \{ P, P_T \}_{DB} = \dot{B} \left( \frac{\partial \phi}{\partial Q} \right).
\end{equation}

From these relations, we can conclude that, although the transformation (\ref{PSCTrans}) is canonical in the extended phase space, it does not preserve the Dirac brackets. Therefore, the Hamilton equations for the physical degrees of freedom $q$ and $p$ are different from the Hamilton equations for the degrees of freedom $Q$ and $P$. In others words, the canonical transformation (\ref{PSCTrans}) in the extended phase space is not canonical when it is restricted to the reduced phase space given by the physical degrees of freedom.

Let us conclude this section by calculating the Generating Function $F= \tilde{q}\, p + \tilde{t}\, p_{\tilde{t}} + F_3(p,p_{\tilde{t}}, Q,T)$ related with this canonical transformation. In this case, $F_3$ is now given by
 \begin{equation} \label{Gfunction1}
F_3(p, p_{\tilde{t}}, Q, T) = - B(T) \, p_{\tilde{t}} - A(Q,T)\, p + \int A'(Q,T)\, D(Q,T) \, dQ,
\end{equation}
\noindent where $B(T)$, $D(Q,T)$, and $A(Q,T)$ are given in (\ref{ExBandD1}), (\ref{ExBandD2}), and (\ref{AinQ}) and (\ref{EqforA}), respectively. This Generating Function will be used in the path integral analysis of the next section.
 
To summarize, we begin with the standard description of the system in the phase space with coordinates $(q,p)$. Due to the Hamiltonian $H(q,p,t)$ being time-dependent, we enlarge the phase space by adding two additional degrees of freedom: $\tilde{t}$ and $p_{\tilde{t}}$. This enlarging of the phase space allows us to consider a canonical transformation that is not only time-dependent but it is also $p_{\tilde{t}}$-dependent. The aim is to ``transfer'' the time-dependence of the Hamilton equations to the coefficients of this canonical transformation, rendering the final Hamilton equations time-independent.  As a result, we obtain three auxiliary equations given in (\ref{EnergyCL}). The first two of them can be easily solved, while the third gives rise to the Equation (\ref{EqforA}), which is a generalization of the auxiliary Equation (\ref{AuxLewis}). The final outcomes of this procedure are the Hamilton equations given by (\ref{NewHEq}). 

A simplified diagram of our procedure is given below
$$ \begin{array}{ccc} \boxed{\mbox{Extended Phase Space} \; (q,p,t,p_t)} & \xLongleftrightarrow{\mbox{Canonical Transformation}} & \boxed{\mbox{Extended Phase Space} \; (Q,P, T, P_T)} \\ \big\uparrow & & \big\downarrow \\  \boxed{\mbox{Reduced Phase Space} \; (q,p) } & \xLongrightarrow{\mbox{Struckmeier-Riedel Transformations }} & \boxed{\mbox{Reduced Phase Space}\; (Q,P)} \end{array} $$

The main result of this section is the derivation of the boundary term in (\ref{NewS}) together with the gauge-invariant observable $I(Q,P)$ given in (\ref{NEWHamil}). The boundary term is absent in the previous classical analysis given by Struckmeier et al. \cite{struckmeier2001invariants}, and we argue that it plays an important role in the quantization of such systems, as we will see in the next section.

%%%%%%%%%%%%%%%%%%%%%%%%%%%%%%%%%%%%%%%%%%%%%%%%%%%%%%%%%%%%
%%%%%%%%%%%%%%%%%%%%%%%%%%%%%%%%%%%%%%%%%%%%%%%%%%%%%%%%%%%%

\section{Path integral analysis} \label{PIA}

In the previous section, we considered the constrained Hamiltonian analysis of the time-dependent Hamiltonian given in (\ref{OldHamilt}) using Dirac's formalism. This section considers the path integral formulation of this system using the action given in (\ref{IniAction}). It is worth mentioning that we can also implement Dirac's quantization procedure; using operators, this way of quantization was implemented in \cite{Fahn}.   In our case, the extended phase space formalism allows us to consider the transformation (\ref{CTEPhaseS}) as a canonical transformation on each of the infinitesimal intervals in which the quantum-extended phase space is split. It will be explained, in this way, that the boundary term given in (\ref{NewS}) is the result of this transformation and that it coincides with the term reported in \cite{chetouani1989generalized}.

Consider the amplitude given by
\begin{equation}
	\langle q_f , t_f | q_i ,t_i \rangle = \int {\cal D}q \, {\cal D}p \, e^{ \frac{i}{\hbar} \int^{t_f}_{t_i} dt \left[ p \dot{q} - H(q,p,t) \right] }\, ,
\end{equation}
\noindent which is the formal expression of the path integral form of the Feynman propagator and where the Hamiltonian $H(q,p,t)$ is given in (\ref{OldHamilt}). The infinitesimal expression of this amplitude takes the form
\begin{equation}
	\langle q_f , t_f | q_i ,t_i \rangle = \lim_{N \rightarrow \infty} \int \prod^{N-1}_{j=1} dq_j \, \prod^{N}_{j=1} \frac{ dp_j }{2 \pi \hbar } e^{  \frac{i}{\hbar} \sum^N_{j=1} \epsilon \left[ p_j \left( \frac{ q_j - q_{j-1}}{\epsilon} \right) -  H\left( \frac{q_j + q_{j-1}}{2}, p_j , \frac{t_j + t_{j-1}}{2}\right) \right] } .\label{AinSPH}
\end{equation}

Note that we are considering this kernel in configuration space, which implies that the coordinates $q's$ are fixed at time $t_i$ and $t_f$ and thus the corresponding product runs from $1$ to $N-1$. On the other hand, since the momenta remain free at the boundaries, their associated product runs from $1$ to $N$.

The expression (\ref{AinSPH}) can be obtained from the path integral formulation in the extended phase space using the amplitude given by
\begin{eqnarray} 
\langle \tilde{q}_f , \tilde{t}_f , \tau_f | \tilde{q}_i, \tilde{t}_i , \tau_i \rangle^{(0)} &=& \int {\cal D}\tilde{q} \, {\cal D}p \, {\cal D}\tilde{t} \, {\cal D}p_{\tilde{t}} \, \delta(\tilde{t} - g(\tau))\, \delta(p_{\tilde{t}} + H) \, e^{ \frac{i}{\hbar} \int^{\tau_f}_{\tau_i} d\tau \left[ p \tilde{q}' + p_{\tilde{t}} \, \tilde{t}'  - \lambda \left( p_{\tilde{t}} + H(  \tilde{q}, p,\tilde{t}) \right) \right] }, \label{PIformal} \\
&=& \lim_{N \rightarrow \infty} \int \prod^{N-1}_{j=1} \delta\left( \tilde{t}_j - g(\tau_j) \right) \, d\tilde{q}_j \, d\tilde{t}_j \prod^{N}_{j=1} \frac{ \delta( p_{\tilde{t}(j)} + H_j) \, dp_j \, dp_{\tilde{t}(j)} }{2 \pi \hbar } \exp\left[ \frac{i}{\hbar} \sum^N_{j=1} \left\{ p_j ( \tilde{q}_j - \tilde{q}_{j-1} ) + \right. \right. \nonumber \\
&&  \left. \left. + p_{\tilde{t}(j)} ( \tilde{t}_j - \tilde{t}_{j-1} ) - \lambda \phi\left( \overline{\tilde{q}}_j, \overline{\tilde{t}}_j, p_j , p_{\tilde{t}(j)} \right) \right\}  \right] , \label{AmpInf}
\end{eqnarray}
\noindent where the symbol ${\langle | \rangle}^{(0)}$ means that Dirac delta functions are inserted in the measure of the path integral. As a result, this amplitude is not equivalent to that of a motion with an arbitrary Hamiltonian function $H( \tilde{q}, p, \tilde{t}, p_{\tilde{t}})$ in the quantum extended phase space, since the physical quantum theory is described in the reduced Hilbert phase space \cite{henneaux1992quantization}. Here, $H_j := H\left( \overline{\tilde{q}}_j, p_j , \overline{\tilde{t}}_j\right)$, $\tilde{q}_j := \tilde{q}(\tau_j)$, $\tilde{t}_j:=\tilde{t}(\tau_j)$ and $\overline{a}_j := \frac{a(\tau_j) + a(\tau_{j-1})}{2}$ for any function $a(\tau)$. The Formula (\ref{AmpInf}) is the infinitesimal expression for the amplitude in the extended quantum phase space.

Note that the function $g(\tau)$ in the Dirac delta gives the time gauge in this system. In Section \ref{EPSForm}, we used the gauge $\tilde{t}, = f(\tau)$, which is equivalent to considering the function $g(\tau) = \tau$. In this section, we use a more general gauge due to, as we proved in Section \ref{CTEPS}, the Lewis invariant  also being a gauge-invariant observable in the extended phase space formalism. In this way, our former derivation should be consistent with the result of the current section.

The infinitesimal expression of the amplitude (\ref{PIformal}) coincides with (\ref{AinSPH}) whenever the Dirac delta functions are evaluated. This asseveration plays an important role in our derivation because it is  the quantum version of the aforementioned phase space enlarging. For this reason, in Appendix \ref{Appendix}, we show that 
\begin{eqnarray}
	\langle \tilde{q}_f , \tilde{t}_f , \tau_f | \tilde{q}_i , \tilde{t}_i , \tau_i \rangle^{(0)} &=& \langle q_f , t_f | q_i ,t_i \rangle, 
\end{eqnarray} 
\noindent whenever $g(\tau)$ satisfies
\begin{equation} \label{CondOnG}
	\tilde{q}(g^{-1}(\tilde{t}_j)) - \tilde{q}(g^{-1}(\tilde{t}_{j-1})) \approx  \frac{dq(t_j)}{dt} \epsilon, \qquad \epsilon := t_{j} - t_{j-1} \, ,
\end{equation}
\noindent where, naturally, the gauge used in the previous section, $g(\tau) = \tau$, fulfills (\ref{CondOnG}), as can be easily checked.

Similarly to the goal of Section \ref{CTEPS}, our aim in this section is to show that under a quantum canonical transformation of the extended phase space variables in the path integral (\ref{PIformal}), the amplitude of the time-dependent Hamiltonian system can be written as the product of two factors. One of the factors is the amplitude of a time-independent Hamiltonian system (related to the invariant $I$ of the previous section),  the other factor is related to the quantum canonical transformation, and the boundary term given in (\ref{NewS}). In order to do so, let us consider the discrete version of the canonical transformation (\ref{CTEPhaseS}). 

In the case of the time scaling, the new time coordinate is related to the old one by $\tilde{t} = B(T)$, as is given in (\ref{CTEPhaseS}); hence, on each interval in (\ref{AmpInf}), we have the relation $\tilde{t}_j \rightarrow T_j$. Accordingly, on each of the intervals $[\tilde{t}_j, \tilde{t}_{j+1}]$, we obtain
\begin{equation}
	\tilde{t}_j = B(T_j), \qquad  \frac{d\tilde{t}_j}{dT_k} =  \dot{B}(T_j) \, \delta_{jk}. \label{TRelation}
\end{equation}
Using (\ref{TRelation}), we obtain the following useful properties for $\bar{\tilde{t}}_j$ and $\Delta \tilde{t}_j$
\begin{equation}
	\bar{\tilde{t}}_j := \frac{\tilde{t}_j + \tilde{t}_{j-1}}{2} \cong B(\overline{T}_j), \qquad \Delta \tilde{t}_j := \tilde{t}_j - \tilde{t}_{j-1} \cong \dot{B}(\overline{T}_j) \Delta T_j  , \label{DTRelation}
\end{equation}
\noindent where $\Delta T_j := T_j - T_{j-1}$ and $\overline{T}_j := (T_j + T_{j-1})/2$. With these relations, an arbitrary function of time $\tilde{t}$, let us say $G(\tilde{t})$, satisfies the following infinitesimal properties 
\begin{equation} \label{CondTD}
 \overline{G}_j := \frac{G(\tilde{t}_j) + G(\tilde{t}_{j-1})}{2} \cong  G(B(\overline{T}_j)), \qquad \Delta G_j := G(\tilde{t}_j) - G(\tilde{t}_{j-1}) \cong \dot{G}(B(\overline{T}_j)) \, \dot{B}(\overline{T}_j) \Delta T_j  .
\end{equation}

In a similar way, the transformations of the coordinate variables $\tilde{q}_j \rightarrow Q_j$ on each interval in (\ref{AmpInf}) are given by
\begin{equation} 
	\tilde{q}_j = \tilde{A}(T_j) \, Q_j , \label{QRelation}
\end{equation}
\noindent and as a result, $\bar{\tilde{q}}_j$ and $\Delta \tilde{q}_j$ take the form
\begin{equation} \label{DQRelation}
\bar{\tilde{q}}_j := \frac{\tilde{q}_j + \tilde{q}_{j-1}}{2} \cong \tilde{A}(\overline{T}_j) \, \overline{Q}_j + \frac{1}{4} \dot{\tilde{A}}(\overline{T}_j) \Delta Q_j \, \Delta T_j, \qquad \Delta \tilde{q}_j := \tilde{q}_j - \tilde{q}_{j-1} \cong \tilde{A}(\overline{T}_j)\,  \Delta Q_j + \dot{\tilde{A}}(\overline{T}_j) \, \overline{Q}_j \Delta T_j,
\end{equation} 
\noindent where $\Delta Q_j := Q_j - Q_{j-1}$ and $\overline{Q}_j := (Q_j + Q_{j-1})/2$.

To derive the transformations for the momentum variables appearing in the measure in (\ref{PIformal}), we have to consider the Generating Function $F_3$ given in (\ref{Gfunction1}). First, note that in the classical analysis, $F_3(p,p_{\tilde{t}}, Q, T)$ provides the following relation between the new and the old momenta
\begin{equation}
	P = - \frac{\partial F_3}{\partial Q}, \qquad P_T = - \frac{\partial F_3}{\partial T}.
\end{equation}
\noindent In the quantum description, the discrete version of these relations provides the relations between both the new and old discrete momenta \cite{chetouani1989generalized} as follows
\begin{equation} 
P_j \cong - \left[ \frac{F_3(p_j, p_{t (j)}, Q_j, \overline{T}_j) - F_3(p_j, p_{t (j)}, Q_{j-1}, \overline{T}_j) }{\Delta Q_j} \right], \quad P_{T(j)} \cong - \left[ \frac{F_3(p_j, p_{t (j)}, \overline{Q}_j, T_j) - F_3(p_j, p_{t (j)}, \overline{Q}_j, T_{j-1}) }{\Delta T_j} \right].
\end{equation} 
\noindent A careful calculation using (\ref{Gfunction1}) results in the explicit expression for the discrete momenta~as
\begin{equation} \label{NMinOM}
P_j =\tilde{A}(\overline{T}_j) p_j - m_0 \left[ \kappa(\overline{Q}_j) + \frac{\dot{\tilde{A}}(\overline{T}_j)}{\tilde{A}(\overline{T}_j)} \overline{Q}_j \right], \quad P_{T\,(j)} = \dot{B}(\overline{T}_j)  p_{t (j)} + \dot{\tilde{A}}(\overline{T}_j) \overline{Q}_j p_j - \frac{m_0 \overline{Q}^2_j}{2} \left[ \frac{\ddot{\tilde{A}}(\overline{T}_j)}{\tilde{A}(\overline{T}_j)} - \frac{\dot{\tilde{A}}^2(\overline{T}_j)}{\tilde{A}^2(\overline{T}_j)}  \right],
\end{equation}
\noindent where the conditions (\ref{CondTD}) were used. We now consider the relations in (\ref{NMinOM}) to express $p_j$ and $p_{t (j)}$ in terms of the momenta $P_j$ and $P_{T(j)}$ as follows
\begin{eqnarray}
p_j &\cong& \frac{P_j }{ \tilde{A}(\overline{T}_j)} + \frac{m_0}{ \tilde{A}(\overline{T}_j) } \left[  \kappa(\overline{Q}_j) + \frac{\dot{\tilde{A}}(\overline{T}_j) \, \overline{Q}_j }{\tilde{A}(\overline{T}_j)}  \right] , \label{pRelation} \\
p_{t(j)} &\cong&  \frac{1}{\dot{B}(\overline{T}_j)} \left\{  P_{T (j)}  - \frac{\dot{\tilde{A}}(\overline{T}_j) \, \overline{Q}_j \, P_j }{ \tilde{A}(\overline{T}_j) } - \frac{m_0  \, \dot{\tilde{A}}(\overline{T}_j) }{ \tilde{A}(\overline{T}_j) } \kappa(\overline{Q}_j) \, \overline{Q}_j + \left[ \frac{\ddot{\tilde{A}}(\overline{T}_j) }{\tilde{A}(\overline{T}_j)} -  \frac{ 3 \dot{\tilde{A}}^2(\overline{T}_j)  }{\tilde{A}^2(\overline{T}_j)}\right] \frac{m_0 \, \overline{Q}^2_j}{2}  \right\} . \label{ptRelation}
\end{eqnarray} 

With these results, we are now ready to analyze the transformation of the measure and the Dirac delta product given in (\ref{PIformal}) or in its infinitesimal version (\ref{AmpInf}). Let us, for simplicity, study the transformation of the measure and the transformation of the Dirac deltas separately. 

In the case of the measure, the calculation yields the following expression 
\begin{eqnarray}
	\prod^{N-1}_{j=1} d\tilde{q}_j \, d\tilde{t}_j \,  \prod^{N}_{j=1} \frac{ dp_{\tilde{t}(j)} dp_j }{2 \pi \hbar } = \prod^{N-1}_{j=1} \tilde{A}(T_j) \, \dot{B}(T_j) \, dQ_j \, dT_j  \prod^N_{j=1} \frac{dP_j\, dP_{T(j)}}{2 \pi \hbar \, \tilde{A}(\overline{T}_j) \, \dot{B}(\overline{T}_j)} , \label{TransM}
\end{eqnarray}
\noindent and note that, in contrast to the analysis given in Chetouani \cite{chetouani1989generalized}, both $d\tilde{t}_j$ and $d p_{\tilde{t} (j)}$ contribute to the change of the measure, the first with a factor $\dot{B}(T_j)$ and the second with a factor $1/ \dot{B}(\overline{T}_j)$.

In the case of the product of the Dirac deltas, we obtain that it transforms as
\begin{eqnarray}
	\prod^{N-1}_{j=1} \delta\left( \tilde{t}_j - g(\tau_j) \right) \prod^{N}_{j=1}  \delta\left( p_{t(j)} + H_j\right) = \prod^{N-1}_{j=1} \frac{\delta(T_j - T^{(0)}_{j})}{  \dot{B}(T^{(0)}_j)  } \prod^N_{j=1}  \dot{B}(\overline{T}_j) \delta(P_{T(j)} -  K_j ) , \label{TransMD}
\end{eqnarray}
\noindent where $T^{(0)}_j$ and $K_j$ are given by
\begin{eqnarray} 
B(T^{(0)}_j) &=& g(\tau_j), \\
K_j &:=& \frac{ \dot{\tilde{A}}(\overline{T}_j) }{\tilde{A}(\overline{T}_j)} \overline{Q}_j \, P_j + m_0 \,  \frac{ \dot{\tilde{A}}(\overline{T}_j) }{\tilde{A}(\overline{T}_j)} \, \overline{Q}_j \, \kappa(\overline{Q}_j) - \left[  \frac{ \ddot{\tilde{A}}(\overline{T}_j) }{\tilde{A}(\overline{T}_j)} -  \frac{ \dot{\tilde{A}}^2(\overline{T}_j) }{\tilde{A}^2(\overline{T}_j)} \right] \frac{m_0 \, \overline{Q}^2_j}{2} - \dot{B}(\overline{T}_j) \, H_j \, . \label{KJDef}
\end{eqnarray}

\noindent Notably, both Dirac deltas contribute with factors $1/\dot{B}(T^{(0)}_j)$ and $\dot{B}(\overline{T}_j)$, respectively. These contributions cancel the factors coming from the differentials $d\tilde{t}_j$ and $d p_{\tilde{t} (j)}$, as we will see further below.

Let us consider the argument of the exponential in (\ref{AmpInf}), which changes under the discrete version of the canonical transformation. Inserting (\ref{TRelation}), (\ref{DTRelation}), (\ref{QRelation}), (\ref{DQRelation}), (\ref{pRelation}) and (\ref{ptRelation}) into the argument of exponential term, we obtain that it takes the following form up to first order in $\Delta Q_j$ and $\Delta T_j$
\begin{eqnarray}
\Phi(P_j, P_{T(j)}, Q_j, Q_{j-1}, T_j, T_{j-1}) &:=& p_j \Delta \tilde{q}_j + p_{t(j)} \Delta \tilde{t}_j \cong P_{T(j)} \,\Delta T_j   +  P_j \, \Delta Q_j  + m_0 \,  \kappa(\overline{T}_j) \, \Delta Q_j  + \nonumber \\
&& + m_0 \frac{ \dot{\tilde{A}}(\overline{T}_j)  }{\tilde{A}(\overline{T}_j)} \overline{Q}_j \, \Delta Q_j + \left[ \frac{ \ddot{\tilde{A}}(\overline{T}_j) }{ \tilde{A}(\overline{T}_j) } - \frac{ \dot{\tilde{A}}^2(\overline{T}_j) }{ \tilde{A}^2(\overline{T}_j) }   \right] \frac{m_0 \, \overline{Q}^2_j \, \Delta T_j}{2} .
\end{eqnarray}

\noindent Of course, due to the classical canonical transformation being a particular type of contact transformation, its quantum analog is also a discrete contact transformation. This is the reason for the linearity of the momenta in this expression. 

We are now ready to calculate the amplitude (\ref{AmpInf}). By inserting the expressions (\ref{TransM}), (\ref{TransMD}), and (\ref{KJDef}) into (\ref{AmpInf}), we obtain
\begin{eqnarray}
\langle \tilde{q}_f , \tilde{t}_f , \tau_f | \tilde{q}_i , \tilde{t}_i , \tau_i \rangle^{(0)} = \lim_{N \rightarrow \infty} \int  \prod^{N-1}_{j=1} \frac{\delta(T_j - T^{(0)}_{j})\,  \tilde{A}(T_j) \, \dot{B}(T_j) \, dQ_j \, dT_j }{  \dot{B}(T^{(0)}_j)  }  \prod^N_{j=1} \frac{\delta(P_{T(j)} -  K_j ) \, dP_j\, dP_{T(j)} \,  e^{ \frac{i}{\hbar} \sum^N_{j=1} \left\{ \Phi - \lambda \phi \right\} } }{2\pi\, \hbar \, \tilde{A}(\overline{T}_j) }  . \label{AmpInfTransf}
\end{eqnarray}
\noindent  Integrating first in the momenta $P_{T(j)}$ and second in the time variables $T_j$, and making use of the Dirac deltas, yields
\begin{equation}
\langle \tilde{q}_f , \tilde{t}_f , \tau_f | \tilde{q}_i , \tilde{t}_i , \tau_i \rangle^{(0)} = \lim_{N \rightarrow \infty} \int  \prod^{N-1}_{j=1} \tilde{A}(T^{(0)}_j) \,  dQ_j  \prod^N_{j=1} \frac{ dP_j  }{2\pi\, \hbar \, \tilde{A}(\overline{T}^{(0)}_j) } e^{ \frac{i}{\hbar} \sum^N_{j=1} \Phi\left(P_j, K_{j}, Q_j, Q_{j-1}, T^{(0)}_j, T^{(0)}_{j-1} \right) } , \label{AmpAF}
\end{equation}
\noindent where 
\begin{eqnarray}
\Phi\left(P_j, K_{j}, Q_j, Q_{j-1}, T^{(0)}_j, T^{(0)}_{j-1} \right) &=& P_j \Delta Q_j - \Delta T_j \left\{  \frac{P^2_j}{2\, m_0} + \kappa(\overline{Q}_j)\, P_j + \dot{B}(\overline{T}_j) \, V + \frac{m_0 \, \kappa^2(\overline{Q}_j)}{2} \right\} + \nonumber \\
&& +  \left[ \kappa(\overline{Q}_j) + \frac{\dot{\tilde{A}}(\overline{T}_j)}{ \tilde{A}(\overline{T}_j) } \overline{Q}_j \right] m_0 \, \Delta Q_j  + \frac{m_0\,\overline{Q}^2_j \, \dot{\tilde{A}}^2(\overline{T}_j) \, \Delta T_j}{2 \, \tilde{A}^2(\overline{T}_j)}.
\end{eqnarray}

Note that the terms in $\dot{B}$ are no longer present in the expression (\ref{AmpAF}). As we mentioned, the contributions of the differentials $d\tilde{t}_j$ and $d p_{\tilde{t} (j)}$ as well as those coming from the Dirac deltas cancel each other. This explains why time reparametrizations cannot modify the measure in the reduced phase space analysis given by Chetouani \cite{chetouani1989generalized}.

To conclude our calculation, we now use the relation
\begin{eqnarray}
4 \, \tilde{A}(T^{(0)}_j)\, \tilde{A}(T^{(0)}_{j-1}) &=& \left( \tilde{A}(T^{(0)}_j) + \tilde{A}(T^{(0)}_{j-1}) \right)^2 - \left( \tilde{A}(T^{(0)}_j) - \tilde{A}(T^{(0)}_{j-1}) \right)^2 \approx  4 \, \tilde{A}(\overline{T}^{(0)}_j)^2 + {\cal O}((\Delta T^{(0)}_j)^2),
\end{eqnarray}

\noindent which implies that, at lowest order in $\Delta T^{(0)}_j$, we have the approximation
\begin{equation}
	\tilde{A}(\overline{T}^{(0)}_j) \cong \sqrt{ \tilde{A}(T^{(0)}_j) \, \tilde{A}(T^{(0)}_{j-1}) }. \label{aprox1}
\end{equation}

By inserting the relation (\ref{aprox1}) into (\ref{AmpAF}), we obtain that the amplitude takes the form
\begin{eqnarray}
\langle \tilde{q}_f , t_f , \tau_f | \tilde{q}_i ,t_i , \tau_i \rangle^{(0)} &=& \frac{1}{\sqrt{\tilde{A}(T_i)\, \tilde{A}(T_f)} } \lim_{N \rightarrow \infty} \int \prod^{N-1}_{j=1} dQ_j \, \prod^{N}_{j=1} \frac{dP_j }{2 \pi \hbar } \,   e^{ \frac{i}{\hbar} \sum^N_{j=1} \left\{ P_j (Q_j -Q_{j-1}) -I_j \, \Delta T_j - {\cal B}_j \right\} } , 
\end{eqnarray}

\noindent where
\begin{eqnarray}
I_j &:=& \frac{P^2_j}{2 \, m_0} + \kappa(\overline{Q}_j) P_j + \overline{V}(\overline{Q}_j) ,\\
{\cal B}_j &:=& m_0 \, \left[ \kappa(\overline{Q}_j) + \frac{\dot{\tilde{A}}(\overline{T}_j) \, \overline{Q}_j}{\tilde{A}(\overline{T}_j)}  \right] \Delta Q_j + \frac{m_0 \, \overline{Q}^2_j }{2} \left[ \frac{\ddot{\tilde{A}}(\overline{T}_j)}{\tilde{A}(\overline{T}_j)} - \frac{ \dot{\tilde{A}}(\overline{T}_j)^2 }{\tilde{A}(\overline{T}_j)^2} \right] \Delta T_j \, .
\end{eqnarray}

It can be seen that the continuum formulation, i.e., the limiting process $N \rightarrow + \infty$,  gives
\begin{eqnarray}
	\langle \tilde{q}_f , t_f , \tau_f | \tilde{q}_i ,t_i , \tau_i \rangle^{(0)} &=&  \frac{ e^{-\frac{i}{\hbar} \left( \int A' D \, dQ \right)\vline^{T_f}_{T_i} }}{\sqrt{\tilde{A}(T_i)\, \tilde{A}(T_f)} }  \int {\cal D}Q \, {\cal D}P\,   e^{ \frac{i}{\hbar} \int^{T_f}_{T_i} \left[  P \frac{dQ}{dT} - I  \right] dT}  .\label{AmpInfTrans}
\end{eqnarray}

This result is a generalization of  that obtained in \cite{chetouani1989generalized}. The difference comes from the fact that we used a generalized canonical transformation given by (\ref{CTEPhaseS}). In this case, we showed that a time reparametrization does not alter the measure of the Feynman propagator: the factors introduced by Dirac deltas transformations cancel the factors introduced by the time reparametrization and the time momentum transformation. In addition to this result, the phase in (\ref{AmpInfTrans}) depends exclusively on the boundary term given in~(\ref{NewS}). This aspect is of course absent in \cite{chetouani1989generalized}.

%%%%%%%%%%%%%%%%%%%%%%%%%%%%%%%%%%%%%%%%%%%
%%%%%%%%%%%%%%%%%%%%%%%%%%%%%%%%%%%%%%%%%%%
%%%%%%%%%%%%%%%%%%%%%%%%%%%%%%%%%%%%%%%%%%%
%%%%%%%%%%%%%%%%%%%%%%%%%%%%%%%%%%%%%%%%%%%
%%%%%%%%%%%%%%%%%%%%%%%%%%%%%%%%%%%%%%%%%%%

\section{Discussion} \label{Discussion}

Plenty of physical systems are modeled with time-dependent Hamiltonians. This  time-dependence makes the study of such systems very difficult . In this realm, the Lewis invariant can be used to solve the equations in both the classical and the quantum descriptions. The Lewis invariant analysis on the extended phase space is of particular importance, and it can be obtained via a canonical transformation in this space. In this work, we applied Dirac's method for constrained systems to time-dependent Hamiltonians and showed that the Lewis invariant is directly a gauge-invariant in the extended phase space. We also showed that the quantum phase relating the Feynman propagator of the time-dependent Hamiltonian $H$ with the Feynman propagator obtained with the Lewis invariant $I$ is given by the boundary term resulting from the canonical transformation of the extended phase~space.

To do so, in Section \ref{EPSForm}, we summarized the main aspects of the Dirac's formalism within the extended phase space with coordinates $(\tilde{q}, \tilde{t}, p, p_{\tilde{t}})$. We obtained the first-class constraint $\phi$ given by~(\ref{Constraint}), which results from the reparametrization invariance of the action (\ref{GAction}). The Dirac brackets are the canonical relations to be used in the canonical quantization of this system and emerge after fixing the gauge with Lagrange multiplier $\lambda =1$. 

In Section \ref{CTEPS}, we studied the canonical transformation (\ref{CTEPhaseS}) which is a generalization of the Struckmeier transformation \cite{struckmeier2002canonical}. We showed that imposing the energy conservation law in the new variables gives rise to the equations (\ref{EnergyCL}). These equations admit solutions in which the mass of the system in the new variables can be considered as a function of the coordinate $Q$, that is, $m_0(Q)$. We restrict this work only to the case in which $m_0$ is constant. This leads to a generalization of the Lewis invariant (\ref{NEWHamil}) obtained by \cite{struckmeier2002canonical}. Moreover, this proves that more general invariants can be attributed to the Hamiltonian $H(q,p,t)$ when the situation $m_0(Q)$ is considered. 

We also derived the Dirac brackets in the variables $Q,T,P,P_T$. These brackets are not canonically related to those in the old variables. This is a direct consequence of the fact that the transformation (\ref{CTEPhaseS}) is canonical in the extended phase space and not in their reduced phase space version. The central relevance of this result appears in the quantum description. Additionally, we obtained the boundary term in the relation (\ref{NewS}), which will emerge as a quantum phase in the Feynman propagator in the last section.

Using these results, we calculated the Feynman propagator for the time-dependent Hamiltonian $H$. We first enlarged the quantum space by considering a measure of the form ${\cal D}\tilde{q}\, {\cal D}\tilde{t}\, {\cal D}p \, {\cal D}p_{\tilde{t}} \, \delta(\tilde{t} - g(\tau)) \, \delta(p_{\tilde{t}} + H)$. With this amplitude, the transformation can be implemented as a canonical map. We obtained that the time reparametrization modifies the measure, but the Dirac deltas' presence absorbs its contributions. As a result, the measure is a gauge-invariant measure, i.e., a time-reparametrization invariant measure. Finally, we obtained that the Feynman propagator (\ref{AmpInfTrans}) is written in terms of the product of two factors. One factor is the usual Feynman propagator in terms of the degrees of freedom $Q, P$ and the Lewis invariant. The other factor is given by
\begin{equation}
	\frac{ e^{-\frac{i}{\hbar} \left( \int A' D \, dQ \right)\vline^{T_f}_{T_i} }}{\sqrt{\tilde{A}(T_i)\, \tilde{A}(T_f)} } ,
\end{equation}  
\noindent where the phase term is clearly the boundary term resulting from the canonical transformation in the extended phase space. The denominator is similar to that reported by Chetouani~\cite{chetouani1989generalized}.

To conclude, let us mention that this work serves as a bridge between two approaches. The first approach is the study of the invariants in the extended phase space using Dirac's formalism, and the second approach is the implementation of the canonical transformations in the quantum extended phase space.  Our work's essence is that it allows the unification of both analyses having as a central aspect  Dirac's formalism and the adequate treatment of the constraint and the degrees of freedom. 

In this spirit, the present result can be expanded, in future work, to more general situations, e.g., cosmology, field theory and quantum gravity. For example, we can apply our results to a gravitational scenario using the Mazur-Mottola formalism \cite{mazur1990path} or to a cosmological one \cite{gueorguiev2020revisiting} and consider a reparametrization procedure like the one introduced by Kucha{\v{r}}~\cite{israel1973relativity}. On the other hand, a similar analysis of the measure but for a relativistic particle with a time-independent mass can be found in \cite{guven1991geometric}. It will be interesting to study if our formalism can also be applied to the case of a relativistic particle with a time-dependent and/or position-dependent mass \cite{alvarez2018relativistic}.

Moreover, recent developments involving the nature of time crystals \cite{wilczek2012quantum, das2018cosmological, shapere2012classical} and quantum speed limits \cite{del2013quantum, lloyd2000ultimate, shanahan2018quantum} arise as another scenario in which our approach might be applied. These last phenomena will be the subject of future investigation in the extended phase space.

\section*{Acknowledgements}

Angel Garcia-Chung acknowledges the total support from DGAPA-UNAM fellowship and also thanks Academia de Matemáticas and Colegio de Física, UP, for the support and enthusiasm.  Daniel Gutiérrez-Ruiz is supported by a CONACyT Ph.D. scholarship No. 332577. This work was partially supported by DGAPA-PAPIIT Grant No. IN103919.

\section{Appendix} \label{Appendix}

Let us consider the differential form of (\ref{PIformal}) given in (\ref{AmpInf}) and let us integrate the $p_{\tilde{t} (j)}$ variables making use of the Dirac delta in the momentum $p_{\tilde{t}}$. These Dirac deltas remove the $p_{\tilde{t} (j)}$ variables, and the expression (\ref{AmpInf}) takes the following form
\begin{eqnarray}
\langle \tilde{q}_f , \tilde{t}_f , \tau_f | \tilde{q}_i , \tilde{t}_i , \tau_i \rangle^{(0)} &=& \lim_{N \rightarrow \infty} \int \prod^{N-1}_{j=1} d\tilde{q}_j \, \prod^{N}_{j=1} \frac{ dp_j }{2 \pi \hbar } \prod^{N-1}_{j=1} d\tilde{t}_j \, \delta\left( \tilde{t}_j - g(\tau_j) \right) e^{ \frac{i}{\hbar} \sum^N_{j=1} \left\{ p_j \left[ \tilde{q}(\tau_j) - \tilde{q}(\tau_{j-1}) \right] - H_j \left[ \tilde{t}(\tau_j) - \tilde{t}(\tau_{j-1}) \right]  \right\} }, \nonumber 
\end{eqnarray}
\noindent where the discrete Hamiltonian $H_j$ is now in the argument of the exponential. We now integrate the time variables $\tilde{t}_j$, and in this case, the Dirac delta in $\tilde{t}_j$ is used. This results in the expression 
\begin{eqnarray}
\langle \tilde{q}_f , \tilde{t}_f , \tau_f | \tilde{q}_i , \tilde{t}_i , \tau_i \rangle^{(0)} &=& \lim_{N \rightarrow \infty} \int \prod^{N-1}_{j=1} d\tilde{q}_j \, \prod^{N}_{j=1} \frac{ dp_j }{2 \pi \hbar }  e^{ \frac{i}{\hbar} \sum^N_{j=1} \left\{ p_j \left[ \tilde{q}(g^{-1}(\tilde{t}_j)) - \tilde{q}(g^{-1}(\tilde{t}_{j-1})) \right] - H_j \left[ \tilde{t}_j - \tilde{t}_{j-1} \right]  \right\} } . \label{AmpInf1}
\end{eqnarray}

The right hand side of this expression is $\tau$-independent if we fix as initial and final times $t_f \equiv \tilde{t}_f = g(\tau_f)$ and $t_i \equiv \tilde{t}_i = g(\tau_i)$. Moreover, let us assume that the following relation $\tilde{q}(\tau) = \tilde{q}(g^{-1}(\tilde{t})) = q(t) $ holds on each of the intervals and that any infinitesimal amount $\delta \tau_j$ induces in the function $g(\tau)$ an infinitesimal amount $\delta t_j $ such that
\begin{equation} 
	\tilde{q}(g^{-1}(\tilde{t}_j)) - q(g^{-1}(\tilde{t}_{j-1})) \approx  \frac{dq(t_j)}{dt} \epsilon, \qquad \epsilon := t_{j} - t_{j-1} \, .
\end{equation}

After imposing the condition (\ref{CondOnG}) on $g(\tau)$, we obtain that the amplitude (\ref{AmpInf1}) has its final form given by (\ref{AinSPH}). That is to say, we show that
\begin{eqnarray}
	\langle \tilde{q}_f , \tilde{t}_f , \tau_f | \tilde{q}_i , \tilde{t}_i , \tau_i \rangle^{(0)} &=& \langle q_f , t_f | q_i ,t_i \rangle, 
\end{eqnarray} 
\noindent whenever $g(\tau)$ satisfies (\ref{CondOnG}). Naturally, the gauge used in Section \ref{EPSForm}, $g(\tau) = \tau$, satisfies this condition, as can be easily checked.

\bibliographystyle{ieeetr}

%%%%%%%%%%%%%%%%%%%%%%%%%%%%%%%%%%%%%%%%%%%
%%%%%%%%%%%%%%%%%%%%%%%%%%%%%%%%%%%%%%%%%%%
%%%%%%%%%%%%%%%%%%%%%%%%%%%%%%%%%%%%%%%%%%%

\end{document}